# Discovery of Late Triassic volcanic ash layers in the deep-water zone of the Nanpanjiang Basin (South China) and the possibility of Carnian Pluvial Episode correlation

Liangjun Wu[a,b,c*]

a. Institute of Karst Geology CAGS/Key Laboratory of Karst Dynamics, MNR & GZAR/International Research Centre on Karst under the Auspices of UNESCO, 541004, Guilin, China

b. Pingguo Guangxi, Karst Ecosystem, National Observation and Research Station, 531406, Pingguo, China

c. College of Earth Sciences, Jilin University, 130061, Changchun, China

The Nanpanjiang Basin is a major basin on the eastern margin of the Tethys Ocean (Fig. 1a). The Late Triassic was an important transition period for the Nanpanjiang Basin changing from marine to continental facies. The shift in the South China region from an extensional platform-basin environment to a foreland basin environment may have occurred during this period (Sun et al., 2016). Simultaneously, climates abrupt change from arid to humid occurred in the Late Triassic (Simms and Ruffell, 1989), known as the Carnian Pluvial Episode (CPE). This event caused an interruption in carbonate sedimentation in global marine platform facies regions, and areas around the Tethys Ocean generally transitioned to fluvial-deltaic sedimentation, such as the Formation of the largest delta system to date in the Boreal Ocean (Klausen and Nyberg, 2019).

Previous research on the CPE in the Nanpanjiang Basin mainly focused on the platform facies of the Zhuganpo Formation and Wayao Formation (Sun et al., 2016; Dal et al., 2024), primarily using biostratigraphy to constrain the timing. However, due to the complexity of Carnian conodont biostratigraphy, no consensus has been reached on a single biostratigraphic scheme (Korte et al., 2003; Orchard, 2010). Moreover, the distribution ranges of most Late Triassic conodont species are long, and the ranges of index fossils such as conodonts and ammonites have not yet met the needs for marine biostratigraphic correlation (Zhang et al., 2018). Therefore, correlation is difficult in the slope-basin zone due to the chronological framework.

The slope-basin area of the Nanpanjiang Basin develops a set of Niluo Member (NL Member) slump-folded nodular limestone, thin-bedded mudstone, and shale (Fig. 1b). Its lithology is significantly different from the turbidites of the overlying and underlying Xuman Formation (XM Formation) and Bianyang Formation (BY Formation), making it a regional marker bed. Regarding the age of the NL Member, one view considers it a marker separating the Anisian and Ladinian stages, being a synchronous heterotopic equivalent of the second Member "lacustrine-swamp layer" of the Yangliujing Formation on the platform (Guizhou Geological Survey, 2013). However, because the NL Member and BY Formation have never had absolute age reports, this age constraint is not definitive. This has led some scholars to believe that no Late Triassic marine strata have been discovered in the Nanpanjiang Basin to date (Yang et al., 2020).

This study found volcanic ash layers within the NL Member in the Wangmo area, Guizhou. This discovery consists of two sections with similar lithology (WM-a and WM-b) (Fig. 1c). Both WM-a and WM-b sections exhibit comparable freshness. Using the nodular limestone of the NL Member and the conglomerate at the base of the BY Formation as marker beds, the correlation between the two sections can be accurately constrained (Fig. 2a, 2b). In the study area, the lithology of the NL Member is consistent with that of the Ziyun-Wangmo region (edge of the slope of the Nanpanjiang Basin), characterized by dark-colored rock series. However, some thick turbiditic sandstone intercalations occur in the upper part of the NL Member. Within the NL

Member, slump folds occur in the nodular limestone, accompanied by abundant pyrite. This indicates deposition in an oxygen-deficient deep-water environment. The underlying XM Formation and overlying BY Formation both exhibit turbidites with Bouma sequences, confirming the NL Member represents a distinct segment within the several-kilometer-thick stratigraphic succession.

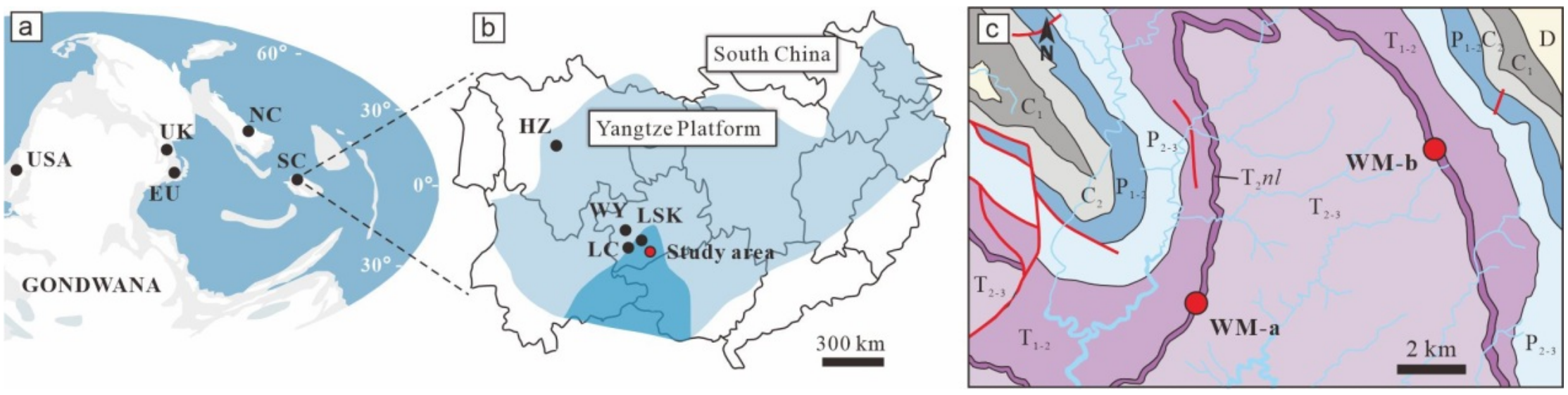

**Fig. 1.** Location of the study area in the Nanpanjiang Basin.
(a) Middle-Late Triassic plate configuration (modified after Scotese, 2016). SC-South China, NC-North China, UK-United Kingdom, USA-United States of America, EU-European Union; (b) Triassic palaeogeographic reconstruction of South China (modified after Sun et al., 2016). HZ-Hanzeng, WY-Wayao, LSK-Laishike, LC-Longchang; (c) Geological map of the study area.

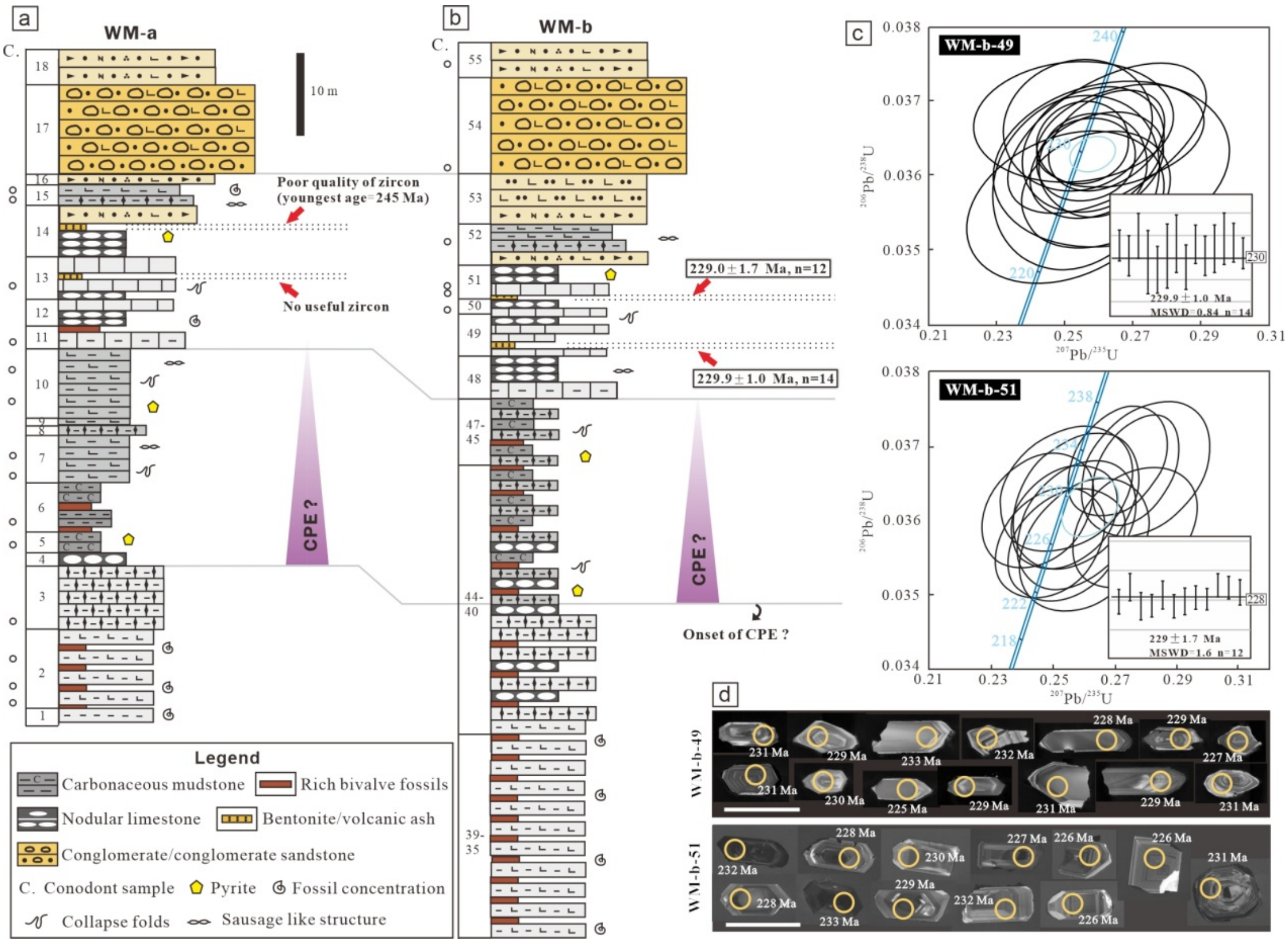

**Fig. 2.** Zircon U-Pb geochronology of volcanic ash from the NL Member, Nanpanjiang Basin.
(a) Stratigraphic column of section WM-a; (b) Stratigraphic column of section WM-b; (c) Concordia plots and weighted mean ages for zircons from volcanic ash WM-b-49 and WM-b-51; (d) Representative cathodoluminescence images of zircons from the two volcanic ash layers. The scale bars in the images are 100 μm.

Both sections developed two layers of volcanic ash (Supplementary Materials Fig. S1). To avoid accidental errors, we sampled twice and sent the samples to Nanjing Hongchuang Exploration Technology Service Co., Ltd and Wuhan Sample Solution Analytical Technology Co., Ltd. for U-Pb dating of zircons respectively (Supplementary Materials Table. S1, Table S2). The results show that the volcanic ash in WM-a section (layers 13 and 14) contained few zircons, mostly fragmented, with the youngest single zircon age at 245 Ma. The zircons from the two volcanic ash layers in the WM-b section were more complete with clear zoning. This indicates that these zircons are of magmatic origin. Layer 49 yielded a weighted age of 229.9±1.0 Ma (MSWD=0.84) (Fig. 2c, 2d), with an average zircon Th/U ratio of 0.417. Layer 51 yielded a weighted age of 229.0±1.7 Ma (MSWD=1.6) (Fig. 2c, 2d), with an average zircon Th/U ratio of 0.775. Both volcanic ash layers indicate that the NL Member is within the Carnian Stage. This age is much younger than the previously suggested depositional time of the NL Member (Anisian/Ladinian boundary, ~241.46 Ma).

In this study, conodont fossil samples were concurrently collected from both sections (sampling locations shown in Fig. 2a, 2b). Unfortunately, none of the 41 conodont samples yielded age-diagnostic fossils. The primary reason may be that prolonged and intense turbiditic environments were unsuitable for conodont survival. Additionally, according to zircon U-Pb dating results, conodont diversity and distribution during this period (Carnian) were far less abundant than in the early Triassic (the results of conodont fossils will be published later). This further validates the difficulty in correlating deep-water and shallow-water chronostratigraphic frameworks using conodonts.

Compared to turbidites, the NL Member represents slow sedimentation (or starved sedimentation). In the deep-water area of the Nanpanjiang Basin, the turbidite sedimentary thickness is enormous: the XM Formation is ＞2700 m, and the BY Formation is ~2120 m (Guizhou Geological Survey, 2013). Although the Triassic was a time of frequent volcanic activity, turbidite sedimentation—due to ample sediment supply—could easily dilute any coeval volcanic ash, making it difficult to preserve. Therefore, the volcanic ash within the NL Member is valuable chronological material. Whereas the upper part of the NL Member contains turbidite interlayers, we estimate that its base is ~234 Ma based on the proportion of slow sedimentation (Supplementary Materials Fig. S2), which coincides with the known initiation age range of the CPE. In shallow water systems, the input of terrigenous clastics or volcaniclastic material is one of the important factors for the demise of platform carbonates (Lehrmann et al., 1998; Wilson and Lokier, 2002). When terrigenous siliciclastic input accelerates into the near-shore basin, suspended matter increases water turbidity, reduces photic zone thickness, inhibits primary productivity, and causes the migration or disappearance of main carbonate deposition (Bodin et al., 2017; Shi et al., 2017). Therefore, the carbonate retreat caused by the CPE might preserve evidence in slope facies environments. The NL Member is distinctly different from the sandstone, siltstone, and shale assemblages in the overlying and underlying strata. It is the only carbonate deposit in the regional slope-basin zones. Moreover, evidence indicates that sedimentation rates increased in the western Tethys during the CPE, while the opposite trend occurred in the eastern Tethys (Dal et al., 2024). Lithofacially, the NL Member matches this characteristic. Therefore, this suite of features suggests that the Niluo Member may record the CPE. These age data serve as a new anchor point for Late Triassic stratigraphic correlation in South China and show potential for correlating CPE sediments.

**Conflict of interest**

The author declares that no conflict of interest.

**Acknowledgments**

This work was supported by the National Natural Science Foundation of China [grant number 42001011]; the Fundamental Research Funds for Central Public Welfare Research Institutes, CAGS [grant number JKYQN202365]; the Guangxi Natural Science Foundation [grant number 2022GXNSFBA035592]; the China Geological Survey [grant number DD20221637].

**Corresponding author**

E-mail address: wuliangjun@mail.cgs.gov.cn (L. Wu).

Supplementary materials for

# Discovery of Late Triassic volcanic ash layers in the deep-water zone of the Nanpanjiang Basin (South China) and the possibility of Carnian Pluvial Episode correlation

Liangjun Wu[a,b,c*]

a. Institute of Karst Geology CAGS/Key Laboratory of Karst Dynamics, MNR & GZAR/International Research Centre on Karst under the Auspices of UNESCO, 541004, Guilin, China

b. Pingguo Guangxi, Karst Ecosystem, National Observation and Research Station, 531406, Pingguo, China

c. College of Earth Sciences, Jilin University, 130061, Changchun, China

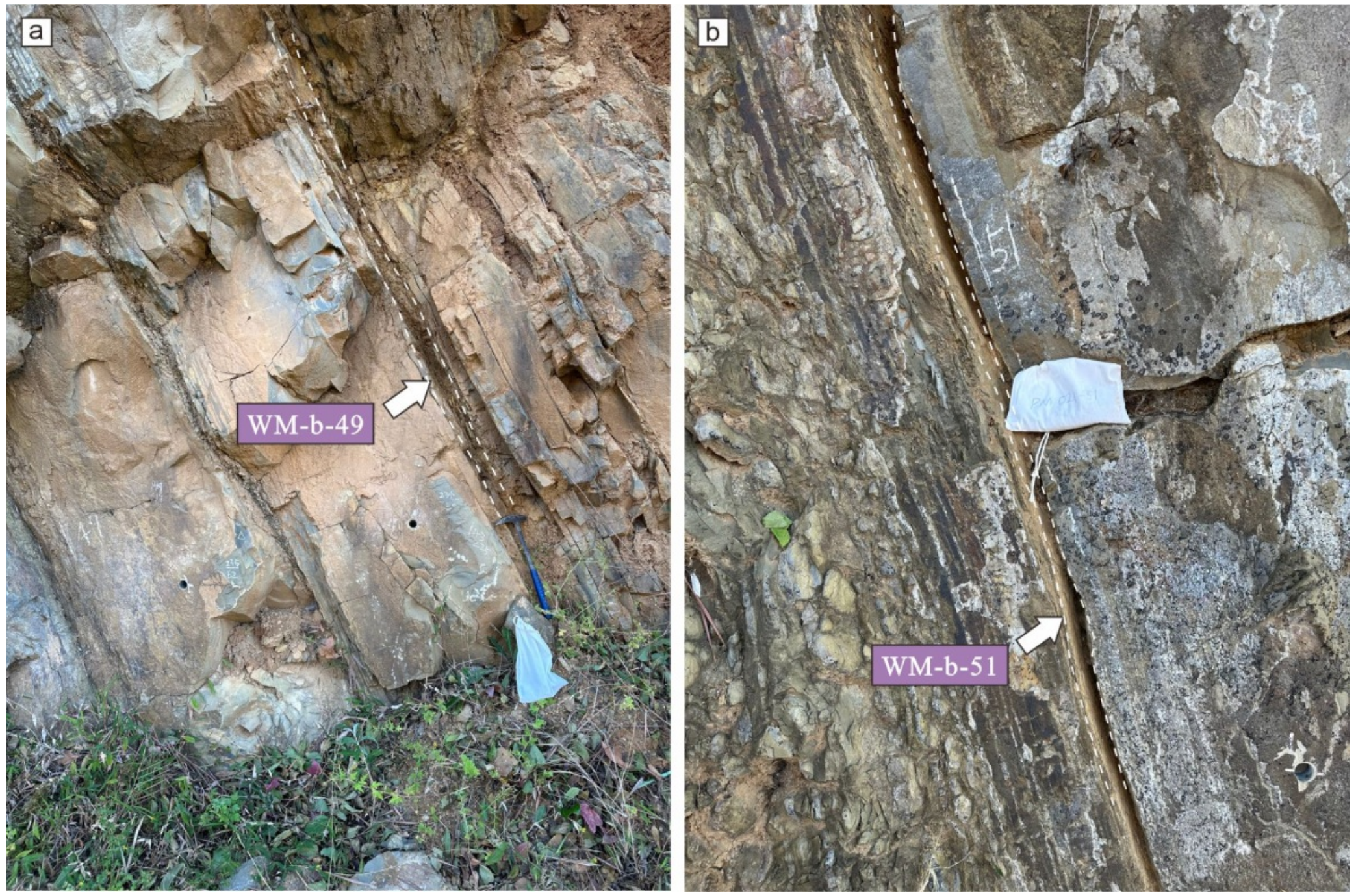

**Fig. S1** The outcrop of volcanic ash layers of the WM-b section

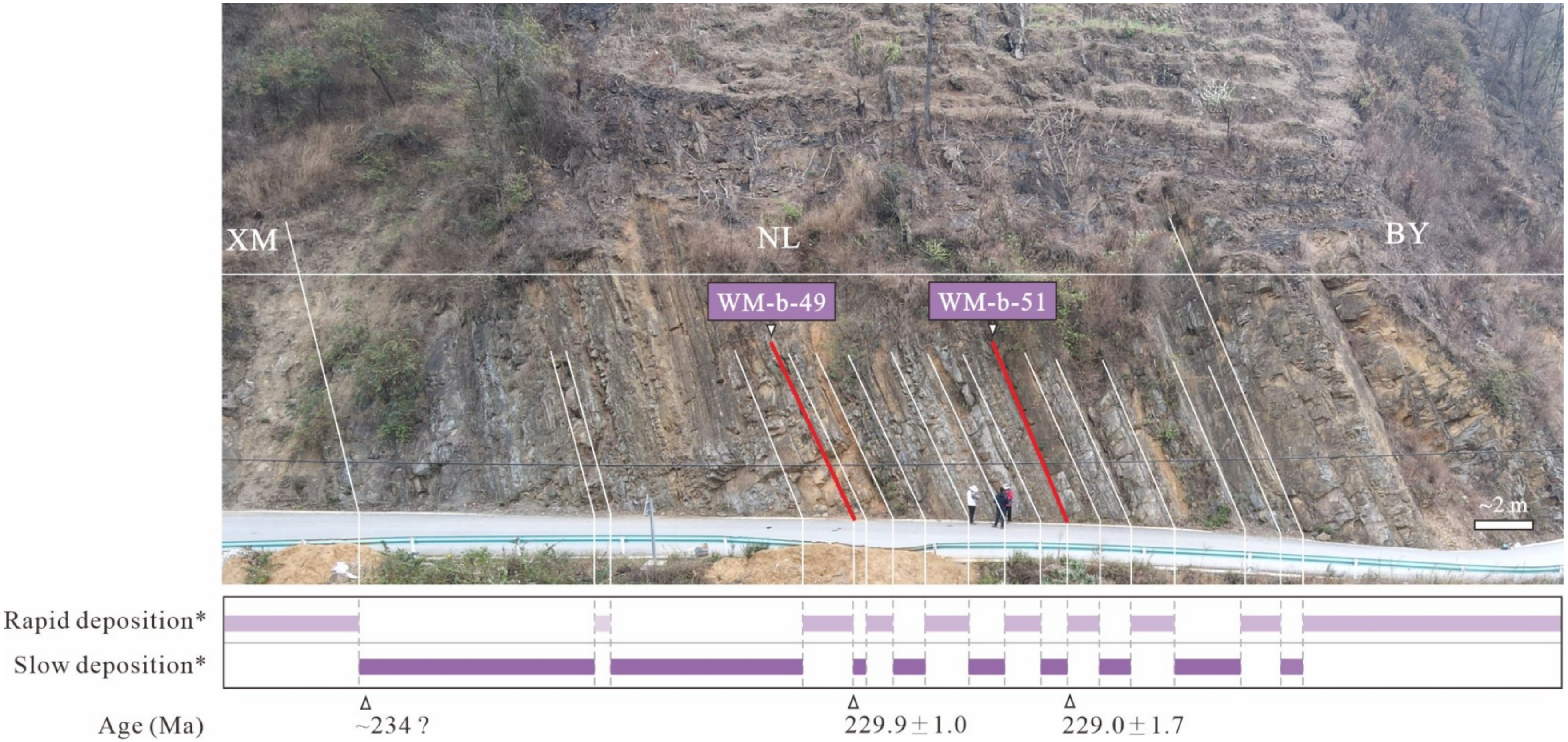

**Fig. S2** Thickness proportion of lithologies formed under rapid and slow sedimentation in the WM-b section

Note: the rapid deposition corresponds to thick-bedded calcareous siltstone and fine sandstone, while the slow deposition corresponds to dark thin-bedded mudstone, shale, and nodular limestone. Due to the wide-angle photography, the scale bar in the figure is approximate. Both the XM member (>2700 m) and BY member (~2120 m) are dominated by extremely rapid turbidite sedimentation, but their sedimentation rates decreased significantly within the CPE [1]. Turbidites occasionally develop in the NL member. Excluding the thickness of rapid deposited turbidites, the onset age of the NL member is estimated to be ~234 Ma based on the thickness of the slow deposited portion.

**Table. S1** LA-ICP-MS zircon U-Pb dating data for the sample of WM-b-49

| Sample | No. | Th | U | Th/U | U-Th-Pb isotope ratio | | | | | | Age (Ma) | | | | | | Concordance |
|---|---|---|---|---|---|---|---|---|---|---|---|---|---|---|---|---|---|
| | | | | | $^{207}$Pb/$^{235}$U | | $^{207}$Pb/$^{206}$Pb | | $^{206}$Pb/$^{238}$U | | $^{207}$Pb/$^{235}$U | | $^{207}$Pb/$^{206}$Pb | | $^{206}$Pb/$^{238}$U | | |
| | | (ppm) | (ppm) | | Ratio | 2σ | Ratio | 2σ | Ratio | 2σ | Age | 2σ | Age | 2σ | Age | 2σ | (%) |
| WM-b-49 | 1 | 217.600 | 11.428 | 0.407 | 0.255000 | 0.011000 | 0.050800 | 0.002300 | 0.036450 | 0.000450 | 229.9 | 9.2 | 220.0 | 100.0 | 230.8 | 2.8 | 100 |
| | 2 | 252.600 | 13.956 | 0.417 | 0.259000 | 0.016000 | 0.052500 | 0.003100 | 0.036150 | 0.000580 | 233.0 | 13.0 | 270.0 | 130.0 | 228.9 | 3.6 | 98 |
| | 3 | 140.200 | 8.091 | 0.412 | 0.252000 | 0.024000 | 0.051100 | 0.005200 | 0.036730 | 0.000660 | 231.0 | 21.0 | 170.0 | 200.0 | 232.5 | 4.1 | 99 |
| | 4 | 175.100 | 9.589 | 0.381 | 0.253000 | 0.020000 | 0.051100 | 0.004000 | 0.035920 | 0.000960 | 228.0 | 17.0 | 220.0 | 170.0 | 227.5 | 6.0 | 100 |
| | 5 | 170.500 | 10.092 | 0.493 | 0.249000 | 0.023000 | 0.050700 | 0.004700 | 0.035580 | 0.000830 | 224.0 | 19.0 | 190.0 | 190.0 | 225.4 | 5.2 | 99 |
| | 6 | 129.700 | 6.167 | 0.343 | 0.264000 | 0.028000 | 0.053200 | 0.005600 | 0.036150 | 0.000930 | 237.0 | 22.0 | 280.0 | 220.0 | 228.9 | 5.8 | 97 |
| | 7 | ~~341.200~~ | ~~47.391~~ | ~~0.049~~ | ~~1.720000~~ | ~~0.120000~~ | ~~0.093000~~ | ~~0.002500~~ | ~~0.133200~~ | ~~0.007300~~ | ~~1006.0~~ | ~~47.0~~ | ~~1479.0~~ | ~~53.0~~ | ~~805.0~~ | ~~42.0~~ | ~~78~~ |
| | 8 | 227.000 | 11.691 | 0.427 | 0.262000 | 0.024000 | 0.051900 | 0.004400 | 0.036510 | 0.000830 | 235.0 | 20.0 | 250.0 | 190.0 | 231.1 | 5.2 | 98 |
| | 9 | 297.900 | 16.943 | 0.419 | 0.259000 | 0.018000 | 0.052800 | 0.003600 | 0.035820 | 0.000640 | 233.0 | 14.0 | 270.0 | 140.0 | 226.9 | 4.0 | 97 |
| | 10 | 333.100 | 16.633 | 0.424 | 0.262000 | 0.014000 | 0.052200 | 0.002700 | 0.036480 | 0.000550 | 236.0 | 11.0 | 270.0 | 120.0 | 231.0 | 3.5 | 98 |
| | 11 | ~~181.100~~ | ~~10.115~~ | ~~0.435~~ | ~~0.272000~~ | ~~0.018000~~ | ~~0.056300~~ | ~~0.003600~~ | ~~0.034960~~ | ~~0.000740~~ | ~~243.0~~ | ~~14.0~~ | ~~400.0~~ | ~~130.0~~ | ~~221.5~~ | ~~4.6~~ | ~~91~~ |
| | 12 | ~~545.400~~ | ~~28.573~~ | ~~0.442~~ | ~~0.275200~~ | ~~0.008500~~ | ~~0.056300~~ | ~~0.001700~~ | ~~0.035390~~ | ~~0.000370~~ | ~~246.5~~ | ~~6.7~~ | ~~445.0~~ | ~~65.0~~ | ~~224.2~~ | ~~2.3~~ | ~~91~~ |
| | 13 | 264.600 | 14.370 | 0.479 | 0.258000 | 0.013000 | 0.051700 | 0.002700 | 0.036150 | 0.000570 | 232.0 | 11.0 | 240.0 | 120.0 | 228.9 | 3.5 | 99 |
| | 14 | 191.900 | 11.542 | 0.464 | 0.258000 | 0.017000 | 0.051400 | 0.003300 | 0.036360 | 0.000690 | 231.0 | 14.0 | 240.0 | 140.0 | 230.2 | 4.3 | 100 |
| | 15 | 302.900 | 15.084 | 0.294 | 0.254000 | 0.018000 | 0.050200 | 0.003600 | 0.036650 | 0.000730 | 229.0 | 14.0 | 180.0 | 160.0 | 232.0 | 4.6 | 99 |
| | 16 | ~~308.800~~ | ~~14.921~~ | ~~0.253~~ | ~~0.268000~~ | ~~0.016000~~ | ~~0.052100~~ | ~~0.003100~~ | ~~0.037110~~ | ~~0.000700~~ | ~~240.0~~ | ~~13.0~~ | ~~270.0~~ | ~~130.0~~ | ~~234.9~~ | ~~4.4~~ | ~~98~~ |
| | 17 | ~~231.700~~ | ~~13.114~~ | ~~0.481~~ | ~~0.309000~~ | ~~0.011000~~ | ~~0.063200~~ | ~~0.002100~~ | ~~0.035250~~ | ~~0.000430~~ | ~~272.4~~ | ~~8.3~~ | ~~675.0~~ | ~~72.0~~ | ~~223.3~~ | ~~2.7~~ | ~~80~~ |
| | 18 | 296.800 | 16.027 | 0.440 | 0.266000 | 0.020000 | 0.052400 | 0.003800 | 0.036540 | 0.000560 | 238.0 | 16.0 | 260.0 | 150.0 | 231.3 | 3.5 | 97 |
| | 19 | ~~802.000~~ | ~~43.003~~ | ~~0.353~~ | ~~0.253000~~ | ~~0.015000~~ | ~~0.051700~~ | ~~0.002800~~ | ~~0.035280~~ | ~~0.000620~~ | ~~228.0~~ | ~~12.0~~ | ~~260.0~~ | ~~120.0~~ | ~~223.5~~ | ~~3.9~~ | ~~98~~ |
| | 20 | 304.900 | 17.476 | 0.429 | 0.256000 | 0.012000 | 0.051500 | 0.002500 | 0.036220 | 0.000450 | 233.0 | 10.0 | 230.0 | 100.0 | 229.4 | 2.8 | 98 |

Brief description of experimental methods and equipments:

The U-Pb dating of sample WM-b-49 was conducted using LA-ICP-MS in Nanjing Hongchuang Exploration Technology Service Co., Ltd. The Resolution SE model laser ablation system (Applied Spectra, USA) was equipped with ATL (ATLEX 300) excimer laser and a Two Volume S155 ablation cell. The laser ablation system was coupled to an Agilent 8900 ICPMS (Agilent, USA).

Detailed tuning parameters can be seen in [2]. LA-ICP-MS tuning was performed using a 50 micron diameter line scan at 3 μm/s on NIST 612 at ~3.5 J/cm$^2$ with repetition rate 10 Hz. Adjusting the gas flow to get the highest sensitivity ($^{238}$U~6*10$^5$ cps) and the lowest oxide ratio(ThO/Th＜0.2%). P/A calibration was conducted on the NIST 610 using a 100 micron diameter line scan. Other laser parameters are the same as that of tuning. Mass analyzed were$^{29}$Si ,$^{31}$P, $^{45}$Sc, $^{49}$Ti, $^{56}$Fe, $^{89}$Y, $^{91}$Zr,$^{93}$Nb, $^{139}$La, $^{140}$Ce, $^{141}$Pr, $^{146}$Nd, $^{147}$Sm, $^{151}$Eu, $^{157}$Gd, $^{159}$Tb, $^{163}$Dy, $^{165}$Ho, $^{166}$Er, $^{169}$Tm, $^{173}$Yb, $^{175}$Lu, $^{178}$Hf, $^{181}$Ta, $^{202}$Hg, $^{204}$Pb, $^{206}$Pb, $^{207}$Pb, $^{208}$Pb, $^{232}$Th, $^{235}$U and $^{238}$U, with a total sweep time of ~ 0.23 seconds. Zircon were mounted in epoxy discs, polished to expose the grains, cleaned ultrasonically in ultrapure water, then cleaned again prior to the analysis using AR grade methanol. Pre-ablation was conducted for each spot analysis using 5 laser shots (~0.3 μm in depth) to remove potential surface contamination. The analysis was performed using 30 μm diameter spot at 5 Hz with fluence of 2.5 J/cm$^2$.

Iolite software package was used for data reduction [3]. Zircon 91500 and GJ-1 was used as primary and secondary reference materials respectively. Triplets of 91500 and GJ-1 were bracketed between multiple groups of 10 to 12 sample unknowns. Typically, 35-40 seconds of the sample signals were acquired after 20 seconds gas background measurement. Using the exponential function to calibrate the downhole fractionation [3]. Iolite software package ('3D Trace Element') was used for data reduction [4]. Measured ages of 91500 (1061.5±3.2 Ma, 2σ) and GJ-1 (601.7±4.2 Ma，2σ) are well within 1% of the accepted age.

**Table. S2** LA-ICP-MS zircon U-Pb dating data for the sample of WM-b-51

| Sample | No. | Th | U | Th/U | U-Th-Pb isotope ratio | | | | | | Age (Ma) | | | | | | Concordance |
|---|---|---|---|---|---|---|---|---|---|---|---|---|---|---|---|---|---|
| | | | | | $^{207}Pb/^{235}U$ | | $^{207}Pb/^{206}Pb$ | | $^{206}Pb/^{238}U$ | | $^{207}Pb/^{235}U$ | | $^{207}Pb/^{206}Pb$ | | $^{206}Pb/^{238}U$ | | |
| | | (ppm) | (ppm) | | Ratio | 1σ | Ratio | 1σ | Ratio | 1σ | Age | 1σ | Age | 1σ | Age | 1σ | (%) |
| WM-b-51 | 1 | 510 | 817 | 0.624207 | 0.2762 | 0.0083 | 0.0558 | 0.0018 | 0.0359 | 0.0003 | 248 | 6.575356 | 456 | 72.215 | 227 | 2.081509 | 91% |
| | 2 | 266 | 399 | 0.665966 | 0.2799 | 0.0085 | 0.0549 | 0.0016 | 0.0367 | 0.0004 | 251 | 6.758405 | 409 | 64.8075 | 232 | 2.322192 | 92% |
| | 3 | 291 | 355 | 0.817975 | 0.2600 | 0.0085 | 0.0528 | 0.0017 | 0.0357 | 0.0004 | 235 | 6.854783 | 320 | 74.065 | 226 | 2.301401 | 96% |
| | 4 | 900 | 714 | 1.260387 | 0.2490 | 0.0071 | 0.0506 | 0.0015 | 0.0357 | 0.0003 | 226 | 5.813204 | 233 | 73.135 | 226 | 1.916309 | 99% |
| | 5 | 254 | 381 | 0.666067 | 0.2573 | 0.0082 | 0.0515 | 0.0016 | 0.0363 | 0.0004 | 233 | 6.663885 | 261 | 72.21 | 230 | 2.360897 | 98% |
| | 6 | 506 | 541 | 0.936755 | 0.2543 | 0.0077 | 0.0517 | 0.0016 | 0.0357 | 0.0003 | 230 | 6.218855 | 272 | 63.88 | 226 | 1.939243 | 98% |
| | 7 | 369 | 541 | 0.681211 | 0.2575 | 0.0078 | 0.0519 | 0.0015 | 0.0359 | 0.0004 | 233 | 6.319014 | 280 | 68.51 | 228 | 2.233965 | 97% |
| | 8 | 583 | 566 | 1.02854 | 0.2450 | 0.0074 | 0.0492 | 0.0015 | 0.0361 | 0.0003 | 223 | 5.999597 | 167 | 74.9875 | 229 | 1.980645 | 97% |
| | 9 | 595 | 551 | 1.079971 | 0.2535 | 0.0070 | 0.0508 | 0.0014 | 0.0361 | 0.0003 | 229 | 5.707914 | 232 | 62.9525 | 228 | 1.798445 | 99% |
| | 10 | 708 | 1244 | 0.569391 | 0.2709 | 0.0065 | 0.0531 | 0.0012 | 0.0368 | 0.0003 | 243 | 5.194293 | 345 | 53.7 | 233 | 1.906205 | 95% |
| | 11 | 299 | 1765 | 0.169653 | 0.2766 | 0.0058 | 0.0546 | 0.0012 | 0.0367 | 0.0003 | 248 | 4.581595 | 394 | 54.625 | 232 | 1.897308 | 93% |
| | 12 | 395 | 493 | 0.801565 | 0.2542 | 0.0087 | 0.0505 | 0.0018 | 0.0364 | 0.0003 | 230 | 7.075693 | 217 | 49.0675 | 231 | 2.16445 | 99% |
| | 13 | ~~242~~ | ~~198~~ | ~~0.624207~~ | ~~0.5782~~ | ~~0.0196~~ | ~~0.0641~~ | ~~0.0023~~ | ~~0.0654~~ | ~~0.0006~~ | ~~463~~ | ~~12.63066~~ | ~~746~~ | ~~41.6625~~ | ~~409~~ | ~~3.381966~~ | ~~87%~~ |
| | 14 | ~~1046~~ | ~~760~~ | ~~0.665966~~ | ~~0.3060~~ | ~~0.0082~~ | ~~0.0606~~ | ~~0.0015~~ | ~~0.0364~~ | ~~0.0003~~ | ~~271~~ | ~~6.374801~~ | ~~626~~ | ~~55.545~~ | ~~231~~ | ~~1.858798~~ | ~~83%~~ |

Brief description of experimental methods and equipments:

This sample WM-b-51, the U-Pb dating and trace element analysis of zircon were simultaneously conducted by LA-ICP-MS at the Wuhan SampleSolution Analytical Technology Co., Ltd., Wuhan, China. Detailed operating conditions for the laser ablation system and the ICP-MS instrument and data reduction are the same as description by [5]. Laser sampling was performed using a GeolasPro laser ablation system that consists of a COMPexPro 102 ArF excimer laser (wavelength of 193 nm and maximum energy of 200 mJ) and a MicroLas optical system. An Agilent 7700e ICP-MS instrument was used to acquire ion-signal intensities. Helium was applied as a carrier gas. Argon was used as the make-up gas and mixed with the carrier gas via a T-connector before entering the ICP. A "wire" signal smoothing device is included in this laser ablation system [6]. The spot size and frequency of the laser were set to ××μm and ××Hz, respectively, in this study. Zircon 91500 and glass NIST610 were used as external standards for U-Pb dating and trace element calibration, respectively. Each analysis incorporated a background acquisition of approximately 20-30 s followed by 50 s of data acquisition from the sample. An Excel-based software ICPMSDataCal was used to perform off-line selection and integration of background and analyzed signals, time-drift correction and quantitative calibration for trace element analysis and U-Pb dating [7,8]. Concordia diagrams and weighted mean calculations were made using Isoplot/Ex_ver3 [9].